\documentclass[journal]{IEEEtran}
\usepackage{cite}

\ifCLASSINFOpdf
  \usepackage[pdftex]{graphicx}
  \graphicspath{{./FIGURES/}}
\else
  \usepackage[dvips]{graphicx}
  \graphicspath{{./FIGURES/}}
\fi
\usepackage[cmex10]{amsmath}
\usepackage{color}

\usepackage{booktabs}
\usepackage{soul}

\begin{document}
%
\title{Exploiting negative differential resistance in monolayer graphene FETs for high voltage gains}
%
%
%

\author{Roberto~Grassi,
        Antonio~Gnudi, \IEEEmembership{Member,~IEEE},
        Valerio~Di~Lecce,
        Elena~Gnani, \IEEEmembership{Member,~IEEE},
        Susanna~Reggiani, \IEEEmembership{Member,~IEEE},
        and Giorgio~Baccarani, \IEEEmembership{Fellow,~IEEE}
\thanks{This work has been supported by the EU project GRADE 317839. The authors acknowledge the CINECA Award N. HP10CNHROP, 2012  for the availability of high performance computing resources and support.}
\thanks{The authors are with the E. De Castro Advanced Research Center on Electronic Systems (ARCES), University of Bologna, 40136 Bologna, Italy (e-mail: rgrassi@arces.unibo.it).}
}

\IEEEpubid{\parbox{\textwidth}{\vspace{0.8cm}\copyright~2013 IEEE. Personal use of this material is permitted. Permission from IEEE must be obtained for all other uses, in any current or future media, including reprinting/republishing this material for advertising or promotional purposes, creating new collective works, for resale or redistribution to servers or lists, or reuse of any copyrighted component of this work in other works.}
}


\maketitle

\begin{abstract}
Through self-consistent quantum transport simulations, we evaluate the RF performance of monolayer graphene FETs in the bias region of negative output differential resistance. We show that, compared to the region of quasi-saturation, a voltage gain larger than 10 can be obtained, at the cost of a decrease in the maximum oscillation frequency of about a factor of 1.5--3 and the need for a careful circuit stabilization.
\end{abstract}

\begin{IEEEkeywords}
Graphene FET, negative differential resistance, terahertz operation, voltage amplifier.
\end{IEEEkeywords}

%
\IEEEpeerreviewmaketitle

\section{Introduction}
\label{sez_intro}

\IEEEPARstart{G}{raphene} has been suggested as a promising material for analog and radio-frequency (RF) applications due to its exceptional electrical properties. In particular, the high mobility and large group velocity can translate to a high device transconductance $g_m$ and high cut-off frequency $f_T$, and there is no need for a band gap to switch off the device as in digital applications \cite{Schwierz10}. Fabricated graphene field-effect transistors (GFETs) exhibiting $f_T$ of hundreds of gigahertz have already been reported \cite{Liao10,Lin10,Wu11,Cheng12}, together with the first applications \cite{Wang10,Lin11,Han11,Guerriero12}. However, challenges still remain. Particularly in short-channel devices, where velocity saturation does not occur, the lack of a band gap leads to poor current saturation (i.e., pronounced drain conductance $g_d$), which negatively affects the device performance as an amplifier. This is especially true at low frequency, where the ability to amplify signals is expressed by the intrinsic voltage gain $g_m/g_d$, which is limited to only few units in monolayer GFETs \cite{Han11,Guerriero12,Wu12NL,Rizzi12}, with a record value of 5.3 for channel lengths of the order of 1~$\mu$m. While voltage gain is not strictly necessary at high frequency, a large $g_d$ also contributes to degrade to some extent the maximum oscillation frequency $f_{max}$, which is the maximum frequency at which power gain can be be obtained, and, in many applications, represents a more important figure of merit than $f_T$ \cite{Schwierz11}.

To address the above issues, the use of bilayer graphene, where a band-gap can be introduced through a vertical electric field, has been suggested and values of $g_m/g_d$ as high as 35 have been experimentally demonstrated \cite{Szafranek12,Fiori12}. In this paper, as an alternative approach, we study whether the RF performance of monolayer GFETs, in particular the voltage gain, can be improved by choosing the bias point in the region of negative differential drain resistance (NDR), i.e., of negative $g_d$. Such NDR has been observed experimentally in long-channel devices \cite{Wu12,Han12} and predicted by numerical simulations for short-channel lengths as well \cite{Dragoman07,NamDo08,Zhao11,Grassi13,Alarcon13,Ganapathi13}, but an analysis of the device small-signal behavior in the NDR region has not been reported yet. Obviously, a device with negative $g_d$ cannot be used as amplifier with a high-impedance load, since the resulting circuit is unstable. However, if the load impedance $1/G_L$ is sufficiently low, the parallel of $g_d$ and $G_L$ can be made positive and, in principle, smaller than the achievable values of $g_d$ in the standard bias region of quasi-saturation, thus potentially resulting in a higher voltage gain.

The objective of the paper is to numerically investigate the feasibility of the above idea. Steady-state simulations are performed to compute the dc $I$--$V$ and $Q$--$V$ characteristics of the device, which allow the extraction of the parameters of a small-signal equivalent circuit and, ultimately, of the analog and RF figures of merit. The device structure, simulation model, and small-signal model are described in Section~\ref{sec_device_model}. The $I$--$V$ characteristics are presented in Section~\ref{sec_dc}, where the connection between $g_m$ and the underlying device physics is also clarified. The circuit stability of the device in the common-source configuration, for a bias point in the NDR region, is analyzed in detail in Section~\ref{sec_stability}. RF performance is discussed in Section~\ref{sec_RF}, followed by conclusions in Section~\ref{sec_conclusions}. 

\IEEEpubidadjcol

\section{Device structure and model}
\label{sec_device_model}

\begin{figure}[!t]
\centering
\hskip0.5cm \includegraphics[scale=0.55]{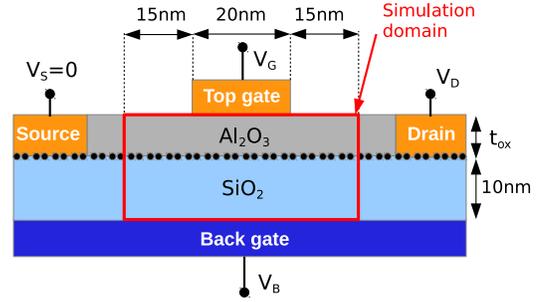}
\caption{Longitudinal cross-section of the device under study. The graphene layer is indicated by the black dots. The function of the back-gate is to dope electrostatically the graphene underlap regions between the top gate and the source and drain contacts. The source voltage is taken as the reference.}
\label{fig_device}
\end{figure}
\begin{figure*}[!t]
\begin{minipage}[b]{0.65\linewidth}
\centering
\includegraphics[height=5cm]{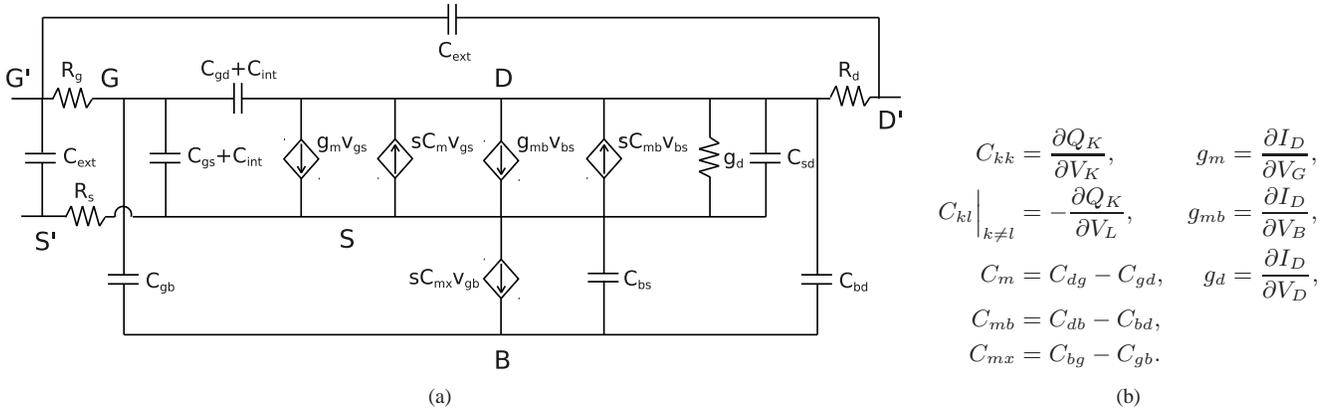}\\
\footnotesize (a)
\end{minipage}
\begin{minipage}[b]{0.35\linewidth}
\centering
\small
\begin{alignat*}{2}
C_{kk} &= \frac{\partial Q_K}{\partial V_K}, &\quad g_m &= \frac{\partial I_D}{\partial V_G},\\
C_{kl}\Big|_{k \ne l} &= -\frac{\partial Q_K}{\partial V_L}, &\quad g_{mb} &= \frac{\partial I_D}{\partial V_B},\\
C_m &= C_{dg} - C_{gd}, &\quad g_d &= \frac{\partial I_D}{\partial V_D},\\
C_{mb} &= C_{db} - C_{bd},& &\\
C_{mx} &= C_{bg} - C_{gb}.& &
\end{alignat*}
\footnotesize (b)
\end{minipage}
\vskip 2mm
\begin{minipage}[c]{\linewidth}
\caption{\label{fig_small_signal}\footnotesize (a) Small-signal equivalent circuit. Here, the unprimed/primed symbols indicate the intrinsic/extrinsic device terminals. (b) Definition of the intrinsic parameters where $k,l \in \{g,d,s,b\}$ \cite{Tsividis99}. $Q_G$ and $Q_B$ are the charges on the top gate and back gate, respectively; $Q_S$ and $Q_D$ are the charges on the graphene layer attributed to the source and drain terminals, respectively.}
\end{minipage}
\end{figure*}
We consider the dual-gate device structure represented in Fig.~\ref{fig_device}, which is similar to the experimental one in \cite{Wu11}, although more aggressively scaled. A gate length $L_g$ of $20$~nm is assumed. The top dieletric layer is Al$_2$O$_3$ ($\kappa=9.5$), while the back dielectric is silicon oxide with thickness of $10$~nm. The top oxide thickness $t_{ox}$ and the back-gate-to-source voltage $V_{BS}$ are treated as parameters. Nominal values are: $t_{ox}=1.2$~nm (effective oxide thickness $\mathrm{EOT}=0.5$~nm) and $V_{BS}=9$~V. Contrary to previous simulations of NDR in GFETs, we do not assume metal-doped \cite{Zhao11,Grassi13} or chemically-doped \cite{Alarcon13,Ganapathi13} source and drain regions, but instead we let the doping of the graphene underlap regions between the top gate and the source and drain contacts be controlled by the back gate. The effective doping corresponding to $V_{BS}=9$~V is about $1.9 \times 10^{13}$~cm$^{-2}$. From a technological point of view, such electrostatic doping technique is easier and more controllable, although the dual gate structure introduces some complications due to the additional wiring and management of the high back-gate voltage.

The simulations are performed using an in-house developed code for GFETs, based on the self-consistent solution of the 2D Poisson equation and the ballistic non-equilibrium Green's function (NEGF) equations \cite{DattaQT2005}, with a $p_z$ tight-binding Hamiltonian and a mode-space solution approach. The model is the same as the one in \cite{Zhao11} but with a different treatment of the interfaces: the source and drain self-energies are computed with the metal-graphene coupling strength $\Delta$ set to zero, and Neumann boundary conditions instead of Dirichlet are used in Poisson's equation at the source and drain ends.
Moreover, instead of assuming a finite channel width with periodic boundary conditions as in \cite{Zhao11}, the device is taken to be infinite in the transverse direction, so that sums over modes are replaced by integrals over the transverse wavevector. The latter are performed by a Gaussian quadrature with $40$ $k$-points.

The small-signal frequency behavior of the device is analyzed through the usual quasi-static approximation, which consists in constructing a small-signal equivalent circuit, whose resistive and capacitive elements are extracted from the dc characteristics of charge and current at the various terminals (Fig.~\ref{fig_small_signal}). The small-signal circuit model is the same used for silicon MOSFETs \cite{Tsividis99}, with the back gate acting as bulk terminal. The source/drain charge $Q_{S/D}$ is taken equal to the charge contribution relative to injection from source/drain of the ballistic transport model. Other charge-partitioning schemes are possible: the authors of \cite{Holland13} have checked that different choices of $Q_S$ and $Q_D$ (with fixed sum $Q_S+Q_D$) have a negligible impact on their results. Source, drain, and gate contact resistances $R_s$, $R_d$, and $R_g$ are included in the model as additional parameters. External parasitic capacitances are modeled according to \mbox{\cite{Koswatta11}}, i.e. through additional capacitive elements between the intrinsic ($C_{int}$) and extrinsic ($C_{ext}$) gate-source and gate-drain pairs of terminals. Unless stated otherwise, we set $R_g$, $C_{int}$, and $C_{ext}$ to zero.

\section{DC characteristics}
\label{sec_dc}

\begin{figure}[!t]
\centering
\includegraphics[scale=0.3]{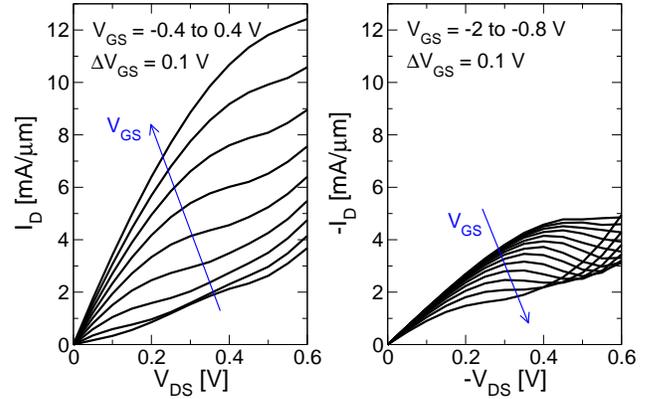}
\caption{Output characteristics for $V_{DS}>0$ and $V_{GS}\ge-0.4$~V, corresponding to an n-type channel (left), and for $V_{DS}<0$ and $V_{GS}\le -0.8$~V, corresponding to a p-type channel (right). At high absolute values of drain bias, quasi-saturation is observed in one case while NDR, i.e. negative drain conductance $g_d$, in the other.}
\label{fig_output}
\end{figure}
The ouput characteristics of the reference device are shown in Fig.~\ref{fig_output}, in two regions of the $V_{DS}$--$V_{GS}$ plane: in the first case ($V_{DS}>0$ and $V_{GS}\ge-0.4$~V) the device operates as an n-type FET and shows quasi-saturation \cite{Meric08,Koswatta11}; in the second case ($V_{DS}<0$ and $V_{GS}\le-0.8$~V), the device operates as a p-type FET and exhibits NDR, confirming previous simulations of GFETs with metal-doped \cite{Zhao11,Grassi13} and chemically-doped \cite{Alarcon13,Ganapathi13} source and drain regions. The different behavior is due to the formation of either an n-n-n or an n-p-n double junction \cite{Grassi13}. The agreement with experiments \cite{Wu12,Han12} is only qualitative due to the gap between the conditions considered in the simulation (20-nm channel length and ballistic transport in ideal graphene) and the limitations of the present graphene technology (e.g., contact resistance and interface effects).

\begin{figure}[!t]
\centering
\includegraphics[scale=0.3]{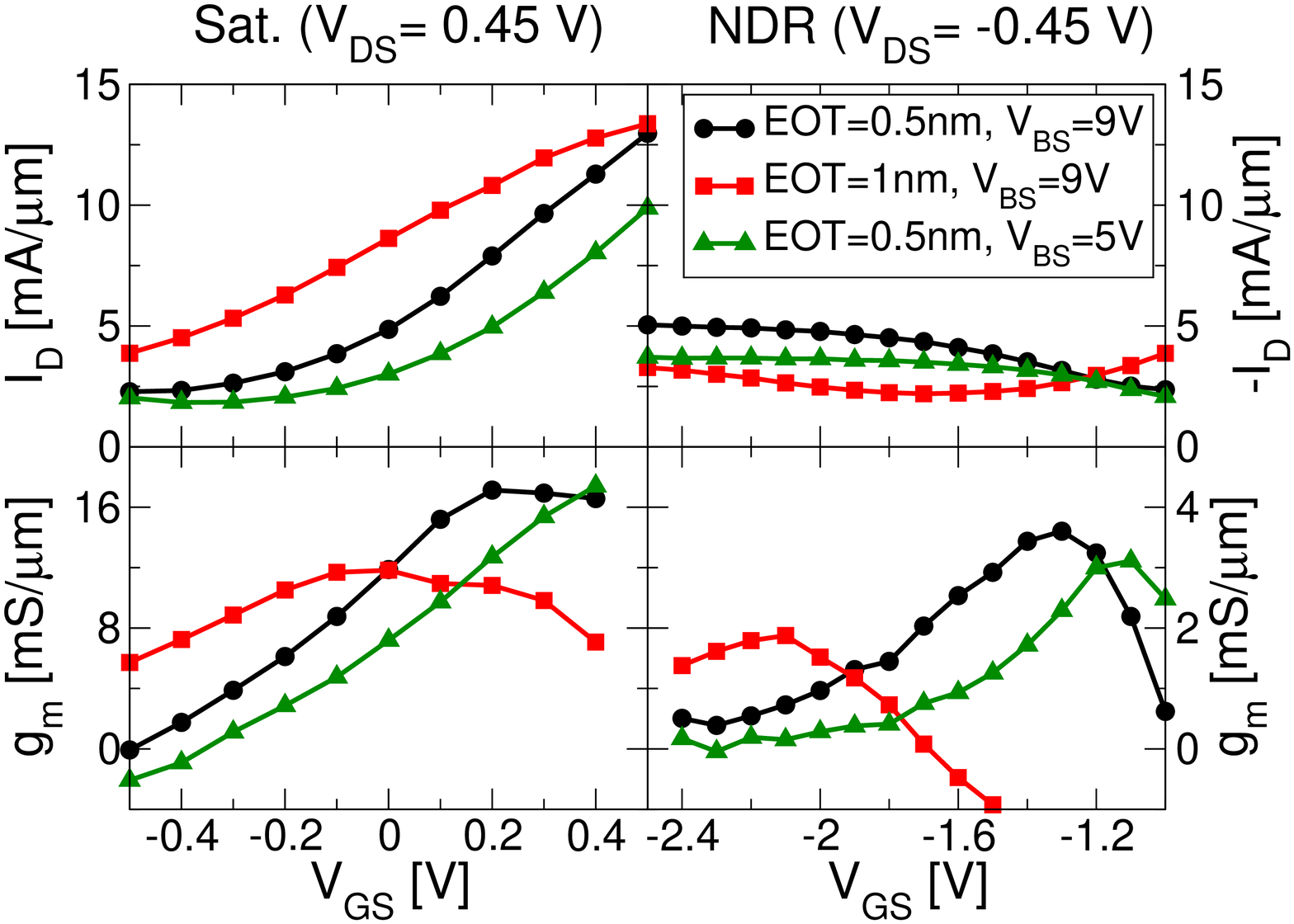}
\caption{Transfer characteristics (top) and corresponding transconductance $g_m$ vs. $V_{GS}$ (bottom) at $V_{DS}=0.45$~V (left) and $V_{DS}=-0.45$~V (right), for the three pairs of EOT and $V_{BS}$ values indicated in the legend.}
\label{fig_transfer}
\end{figure}

NDR is obtained at the cost of lower $I_D$ and transconductance $g_m$, as is evident from the trans-characteristics and the corresponding $g_m$ vs. $V_{GS}$ plots in Fig.~\ref{fig_transfer}. The peak $g_m$ decreases by more than a factor of four. The reason can be ascribed to: \emph{(i)} reduced transmission due to double band-to-band tunneling across the n-p-n junction \cite{Low09}; \emph{(ii)} a transport-mode bottleneck effect induced by the Dirac point at the drain side \cite{Grassi13}. Similar asymmetric performance with respect to top gate bias in dual-gated structures is observed in experiments \cite{Lin10EDL,Zhu13}.
In Fig.~\ref{fig_transfer}, we also show the results obtained by increasing $t_{ox}$ to $2.4$~nm ($\mathrm{EOT}=1$~nm) or by lowering $V_{BS}$ to $5$~V (effective doping of $1 \times 10^{13}$~cm$^{-2}$). As one might expect, a larger EOT leads to significant degradation of the peak $g_m$, in both bias regions, due to reduced electrostatic control of the top gate on the channel potential. In the bias region corresponding to NDR, a lower $V_{BS}$ also goes in the direction of decreasing the peak $g_m$, highlighting the importance of a heavy doping of the source and drain regions in this transport regime.
\begin{table*}[!t]
\renewcommand{\arraystretch}{1.3}
\caption{Intrinsic small-signal parameters at $V_{DS}=0.45$~V and $V_{GS}=0.2$~V (Sat.), and at $V_{DS}=-0.45$~V and $V_{GS}=-1.3$~V (NDR).}
\label{tab_small_signal}
\centering
\begin{tabular}{c*{12}{c}}
\hline
 & $C_{gd}$ & $C_{gs}$ & $C_{sd}$ & $C_{gb}$ & $C_{bd}$ & $C_{bs}$ & $C_{m}$ & $C_{mb}$ & $C_{mx}$ & $g_{m}$ & $g_{mb}$ & $g_d$ \\
 & [aF/$\mu$m] & [aF/$\mu$m] & [aF/$\mu$m] & [aF/$\mu$m] & [aF/$\mu$m] & [aF/$\mu$m] & [aF/$\mu$m] & [aF/$\mu$m] & [aF/$\mu$m] & [mS/$\mu$m] & [mS/$\mu$m] & [mS/$\mu$m] \\
\hline
Sat. & 221 & 565 & 273 & 37 & 50 & 84 & -306 & -9 & -3 & 17.1 & 0.8 & 5.0 \\
NDR & -100 & 985 & 617 & 30 & 42 & 87 & 169 & 8 & 8 & 3.8 & $\simeq$ 0 & -6.1 \\
\hline 
\end{tabular}
\end{table*}
Increasing further the drain degeneracy with respect to the case with $V_{BS}=9$~V would require the use of a high-$\kappa$ substrate material, for the vertical electric field in the back dielectric is already close to the SiO$_2$ limit of $1$~V/nm at $V_{BS}=9$~V. On the other hand, since a higher $\kappa$ also implies a larger back gate capacitance, the back oxide thickness (and consequently $V_{BS}$) should be increased to avoid a counter-productive effect on $g_m$.

The extracted small-signal parameters for the bias points of peak $g_m$ of the quasi-saturation and NDR regions are provided for reference in Table~\ref{tab_small_signal}. Since the $Q$--$V$ characteristics (not shown) were found to be affected by numerical noise, we used a Savitsky-Golay filter of order two \cite{NumRecipes} to compute the parameters in Table~\ref{tab_small_signal}, rather than using finite differences as for the $g_m$ plots in Fig.~\ref{fig_transfer}.

\section{Stability analysis}
\label{sec_stability}

\begin{figure}[!t]
\begin{minipage}[b]{0.6\linewidth}
\centering
\includegraphics[scale=0.4]{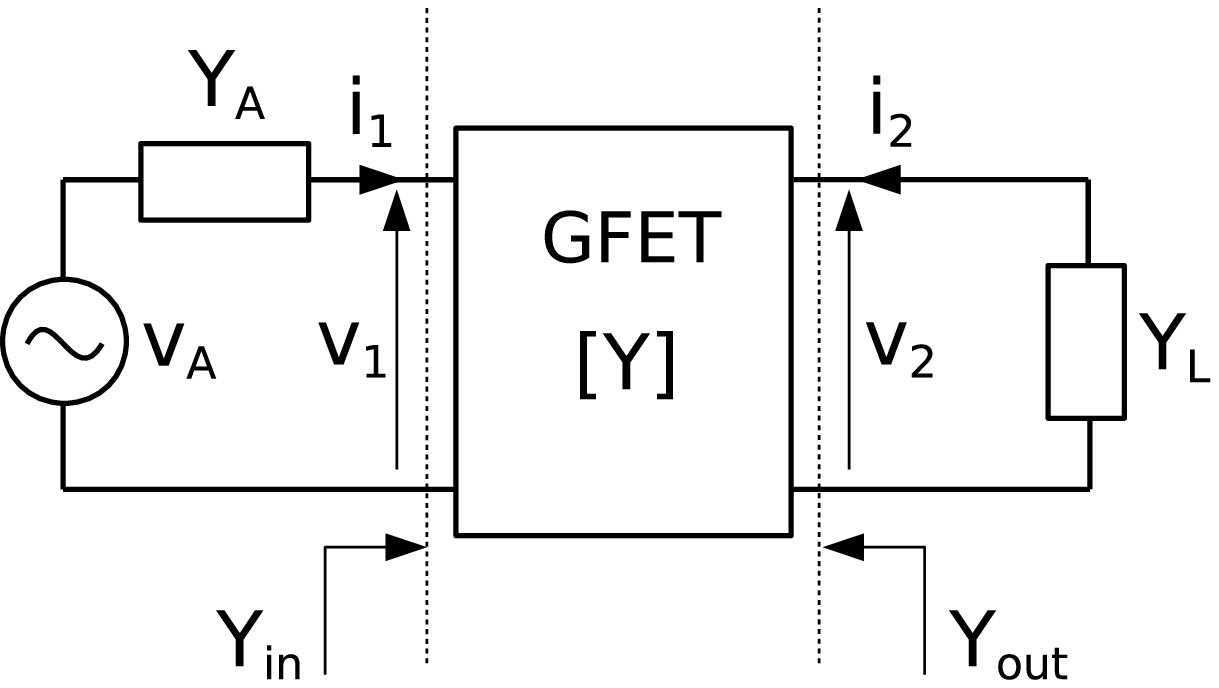}\\
\footnotesize (a)
\end{minipage}
\begin{minipage}[b]{0.18\linewidth}
\centering
\includegraphics[scale=0.4]{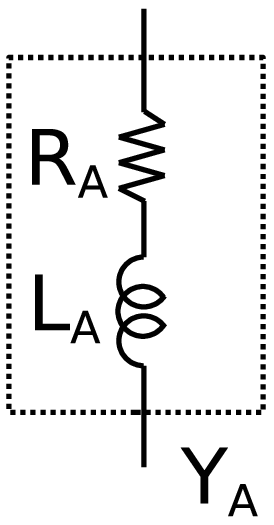}\\
\vskip0.2cm
\footnotesize (b)
\end{minipage}
\begin{minipage}[b]{0.2\linewidth}
\centering
\includegraphics[scale=0.4]{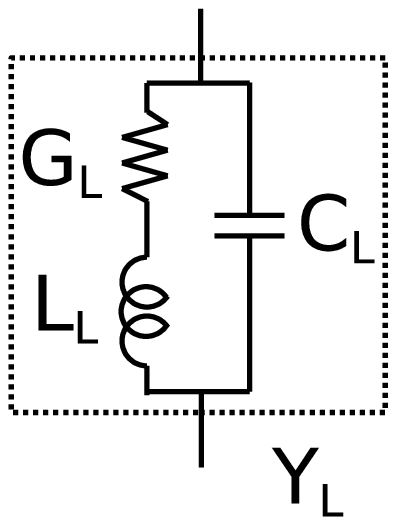}\\
\vskip0.2cm
\footnotesize (c)
\end{minipage}
\caption{(a) Representation in terms of $Y$-parameters of the GFET connected to the source and load networks and symbol definitions. (b-c) Models for the source and load admittances, respectively, which were used to evaluate circuit stability. $L_A$ and $L_L$ represent parasitic inductances.}
\label{fig_circuit}
\end{figure}
In this section, we study the stability of the reference device in the common-source configuration, at $V_{DS}=-0.45$~V and $V_{GS}=-1.3$~V (operating point of peak $g_m$ of the NDR region). We consider a channel width $W=1$~$\mu$m. From the small-signal circuit of Fig.~\ref{fig_small_signal}, one can derive the expressions of the $Y$-parameters of the extrinsic transistor $Y_{11} \ldots Y_{22}$, which allow to compute the output and input admittances $Y_{out}$ and $Y_{in}$ as a function of the source and load admittances $Y_A$ and $Y_L$, respectively (see circuit in Fig.~\ref{fig_circuit}(a) for symbol definitions). Regarding the functional dependence of $Y_A$ and $Y_L$ on frequency $f$, we assume the circuit models in Figs.~\ref{fig_circuit}(b)-(c), where $L_A$ and $L_L$ represent series parasitic interconnect inductances.

\begin{figure}[!t]
\centering
\includegraphics[scale=0.3]{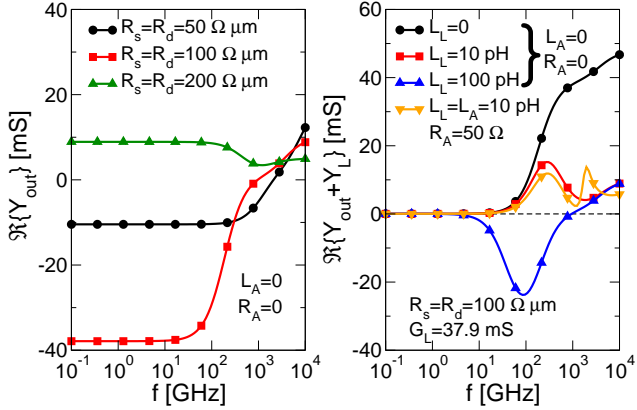}
\caption{\label{fig_stability_a}\footnotesize Stability analysis at the output port for a $1$-$\mu$m-wide device at $V_{DS}=-0.45$~V and $V_{GS}=-1.3$~V (NDR regime). Left: real part of the output admittance $Y_{out}$ as a function of frequency for different values of source/drain contact resistance, assuming short-circuit at the input port. Right: real part of $Y_{out}+Y_L$ vs. frequency for fixed values of contact resistance and $G_L$, and for different $L_L$, $L_A$, and $R_A$, as indicated in the legend. Negative values of $\Re\{Y_{out}+Y_L\}$ indicate circuit instability.}
\end{figure}

The stability of an RF amplifier is usually ensured by requiring that both $Y_{out}$ and $Y_{in}$ have a positive real part in the whole frequency range where the amplifier behaves as an active network \cite{Pozar05}:
\begin{equation} \label{eq_stability1}
\Re\{Y_{out}\}>0,\quad \Re\{Y_{in}\}>0 .
\end{equation}
The ``stability circles'' technique then allows to find on the Smith chart the range of values of $Y_A$ and $Y_L$ for which (\ref{eq_stability1}) are satisfied at each frequency. Such approach, however, cannot be applied in the present case, since $\Re\{Y_{out}\}$ is potentially negative at low frequency, where the device is unilateral. The real part of $Y_{out}$ is plotted as a function of frequency in Fig.~\ref{fig_stability_a}-left for the case of $|Y_A|\equiv\infty$ (short-circuit at the input port) and for values of $R_s=R_d$ from 200 down to 50~$\Omega \cdot \mu$m, i.e. from typical experimental values down to best achievable ones \cite{Moon12}.
The low-frequency value of $Y_{out}$, given by the output conductance $g_{out}$ of the extrinsic transistor
\begin{equation}
g_{out} = \frac{g_d}{1 + (R_s + R_d) g_d + R_s (g_m + g_{mb})} \, , \label{eq_gout}
\end{equation}
is strongly affected by the source and drain contact resistances, as shown in the figure, but is independent of $Y_A$. We note that stability is still possible if
\begin{equation} \label{eq_stability2}
\Re\{Y_{out}+Y_L\}>0,\quad \Re\{Y_{in}+Y_A\}>0,
\end{equation}
which represent less restrictive requirements than (\ref{eq_stability1}). Acceptable values of $Y_A$ and $Y_L$ must satisfy both inequalities simultaneously. Let us first start by assuming $|Y_A|\equiv\infty$. We consider here a value of contact resistance of $100$~$\Omega \cdot \mu m$, for which $g_{out}=-37.9$~mS. To satisfy the first inequality in (\ref{eq_stability2}) at low frequency, the negative value of $g_{out}$ must be compensated by a load conductance $G_L > -g_{out}$, as already mentioned in Section~\ref{sez_intro}. However, a too large parasitic inductance $L_L$ might cancel the effect of $G_L$ at high frequency, as illustrated by the plot of $\Re\{Y_{out}+Y_L\}$ in Fig.~\ref{fig_stability_a}-right for different values of $L_L$ (see only the curves with $L_A=R_A=0$, the other ones being discussed later). We find that $L_L$ must be limited to $\approx10$~pH, an upper bound which should be compatible with an integrated version of the amplifier. Having fixed the values of $G_L$ and $L_L$ this way, we look for values of load capacitance $C_L$, source resistance $R_A$, and $L_A$ that allow to satisfy the second inequality in (\ref{eq_stability2}).
\begin{figure}[!t]
\centering
\includegraphics[scale=0.3]{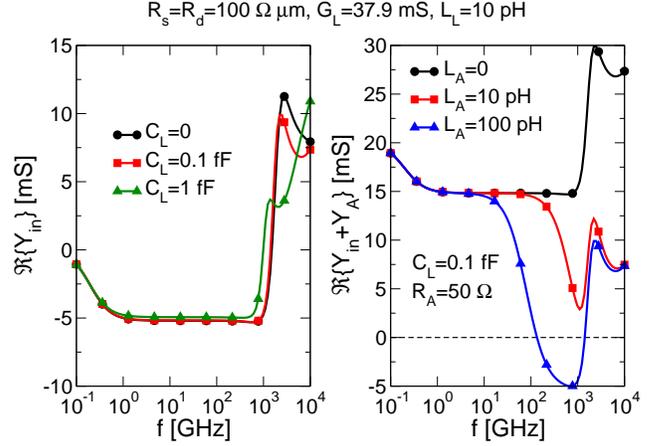}
\caption{Stability analysis at the input port for the same device and bias as in Fig.~\ref{fig_stability_a}, and for $R_s=R_d=100$~$\Omega \cdot \mu$m. $G_L$ and $L_L$ are fixed at values such that $\Re\{Y_{out} + Y_L\}_{|Y_A|\equiv\infty}>0$ (see Fig.~\ref{fig_stability_a}-right). Left: real part of the input admittance $Y_{in}$ as a function of frequency for different values of load capacitance $C_L$. Right: real part of $Y_{in}+Y_A$ vs. frequency for fixed $C_L$ and $R_A$, and for different $L_A$. Negative values of $\Re\{Y_{in}+Y_A\}$ indicate circuit instability.}
\label{fig_stability_b}
\end{figure}
As shown in Fig.~\ref{fig_stability_b}-left, the real part of $Y_{in}$ is not significantly affected by $C_{L}$. Its negative plateau can be compensated by sufficiently low values of $R_A$ and $L_A$ (Fig.~\ref{fig_stability_b}-right). We find that choosing $R_A=50$~$\Omega$, which is the typical characteristic impedance of a transmission line, together with the same upper bound of $10$~pH for $L_A$ as for $L_L$, provides $\Re\{Y_{in} + Y_A\}>0$. The stability of the circuit is finally demonstrated by checking that the same values of $R_A$ and $L_A$ also give $\Re\{Y_{out} + Y_L\}>0$ (triangles down in Fig.~\ref{fig_stability_a}-right). Of course, instead of the procedure outlined here, one could also have tested the circuit stability using the standard pole analysis.

In summary, we have shown that is possibile to ensure the stability of the circuit by canceling out the negative real part of $Y_{out}$ and $Y_{in}$ through a proper choice of the load conductance $G_L$ and the source resistance $R_A$, respectively. The procedure requires: \emph{(i)} an estimate of the contact resistances and hence of $g_{out}$; \emph{(ii)} small enough parasitic interconnect inductances at the input and output port, or the effect of $G_L$ and $R_A$ is made void at high frequency.

\section{Analog and RF metrics}
\label{sec_RF}

The following figures of merit are evaluated for the device in the common-source configuration and biased in either the NDR or quasi-saturation region: dc voltage gain $A_{v0}=v_2/v_1|_{f=0}$ with load $G_L$, cut-off frequency $f_T$, and maximum oscillation frequency $f_{max}$. From the small-signal circuit in Fig.~\ref{fig_small_signal}, a simple expression for $A_{v0}$ can be derived: 
\begin{equation} \label{eq_av0}
A_{v0} = -\frac{g_m/g_d}{1 +G_L/g_{out}}, \\
\end{equation}
where $g_{out}$ is given by (\ref{eq_gout}). $f_{T}$ is obtained by extrapolating the low-frequency short-circuit current gain $|H_{21}|=|Y_{21}/Y_{11}|$ to unity at $-20$~dB/dec, and $f_{max}$ by extrapolating the low-frequency maximum stable gain $\mathrm{MSG}=|Y_{21}/Y_{12}|$ to unity at $-10$~dB/dec. $f_{max}$ is defined here with reference to MSG rather than Mason's unilater gain \cite{Pozar05}, since the latter cannot be defined in the NDR region. In the case of $R_g=C_{int}=C_{ext}=0$, analytical expressions for $f_T$ and $f_{max}$ can be derived by isolating the $1/s$ terms in the expansion of $Y_{21}(s)/Y_{11}(s)$ and $Y_{21}(s)/Y_{12}(s)$, respectively, and by equating their magnitude to unity:
\begin{align}
f_T &= \frac{g_m/( 2 \pi)}{D}, \,\,\, D=\left| C_{gg} \left[ 1 + (R_s + R_d) g_d + R_s g_{mb} \right] + \right. \nonumber \\
&\hspace{2cm} \left. {}+ C_{gd} (R_s + R_d) g_m + C_{gb} R_s g_m \right|, \label{eq_ft} \\
f_{max} &= \frac{g_m/( 2 \pi)}{\left| (C_{gs}+C_{gd}) R_s g_d + C_{gd} \left[ 1 + R_s (g_m + g_{mb}) \right] \right|} , \label{eq_fmax}
\end{align}
where the total gate capacitance $C_{gg}$ is related to the circuit elements in Fig.~\ref{fig_small_signal} through $C_{gg} = C_{gs} + C_{gd} + C_{gb}$.

Let us start considering the peak-$g_m$ bias point of the NDR region ($V_{DS}=-0.45$~V, $V_{GS}=-1.3$~V) and a value of contact resistance $R_s=R_d=100$~$\Omega \cdot \mu$m. Again, we assume $W=1$~$\mu$m.
\begin{figure}[!t]
\centering
\includegraphics[scale=0.3]{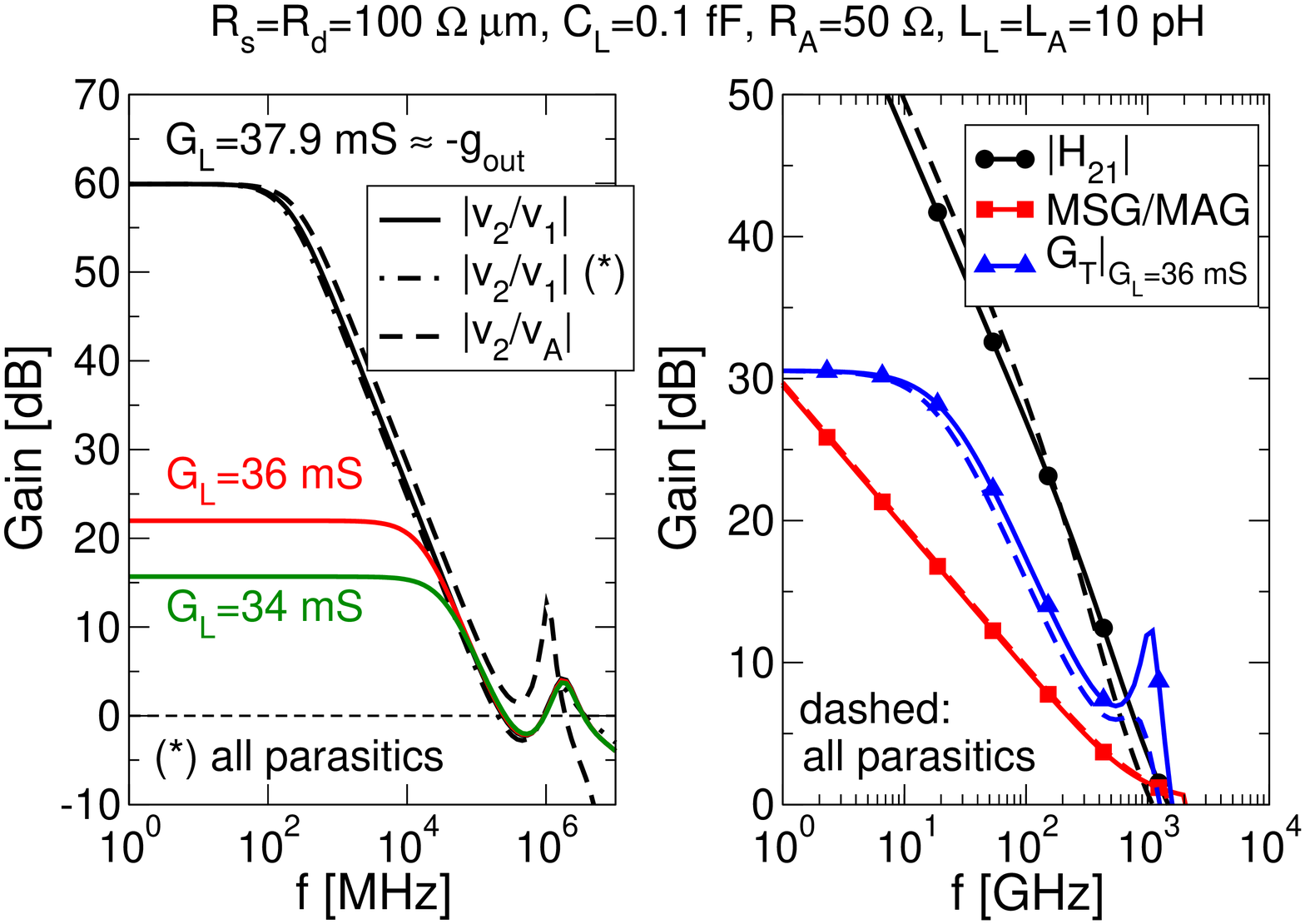}
\caption{Plots of voltage gains $|v_2/v_1|$ and $|v_2/v_A|$ vs. frequency (left) and of short-circuit current gain $|H_{21}|$, MSG or MAG (depending on frequency), and transducer gain $G_T$ vs. frequency (right). The dot-dashed curve in (left) and the dashed curves in (right) are obtained with $R_g=4$~$\Omega$ and $C_{int}=C_{ext}=0.1$~fF. Device width and bias are the same as in Fig.~\ref{fig_stability_a}--\ref{fig_stability_b}. The other parameters are indicated in the legend.}
\label{fig_gain_vs_f}
\end{figure}
Fig.~\ref{fig_gain_vs_f}-left shows the frequency magnitude response of the voltage gain $v_2/v_1$ for different values of $G_L$ (and fixed values of $C_L$ and $L_L$). In accordance with (\ref{eq_av0}), the low-frequency value $|A_{v0}|$ is strongly peaked around $G_L=-g_{out} \approx 37.9$~mS. If the difference between $G_L$ and $-g_{out}$ is less than $5\%$, a voltage gain larger than 10 can be obtained. Furthermore, the larger the gain, the smaller the corresponding bandwidth, resulting in an approximately constant gain-bandwidth product GBW of about $200$~GHz. In the same figure, we also plot the frequency response of $|v_2/v_A|$ for $G_L=37.9$~mS (see legend for values of $R_A$ and $L_A$), which is found to be almost identical to $|v_2/v_1|$, indicating a minor effect of the source admittance. Even when including the additional parasitics $R_g$, $C_{int}$, and $C_{ext}$, the frequency response does not change significantly. The value of $R_g=4$ $\Omega$ considered here has been calculated in a similar way to \mbox{\cite{Holland13}}, by assuming a tungsten gate (resistivity of 56~n$\Omega\cdot$m) of dimensions $W \times L_g \times t_g = 1$~$\mu$m $\times$ $20$~nm $\times$ $60$~nm, contacted on both sides \mbox{\cite[Eq.~9.6.2]{Tsividis99}}. The value of $C_{int}=C_{ext}=0.1$~fF is the same as in \mbox{\cite{Koswatta11}}. In Fig.~\ref{fig_gain_vs_f}-right the frequency response of the current gain $|H_{21}|$ and of MSG is reported. For high enough frequencies, where the transistor becomes unconditionally stable, MSG is replaced by the maximum available gain MAG \cite{Pozar05} (almost invisible in Fig.~\ref{fig_gain_vs_f}-right). In the case of $R_g=C_{int}=C_{ext}=0$, values of $f_T=2.3$~THz and $f_{max}=890$~GHz are extracted. For comparison purposes, we also plot the transducer power gain $G_T$, obtained with source and load parameters (see legend) that ensure the stability of the amplifier. It can be seen that $G_T$ falls off to one at a frequency not too far from $f_{max}$. Again, the inclusion of additional parasitics has only a limited impact.

\begin{figure}[!t]
\centering
\includegraphics[scale=0.3]{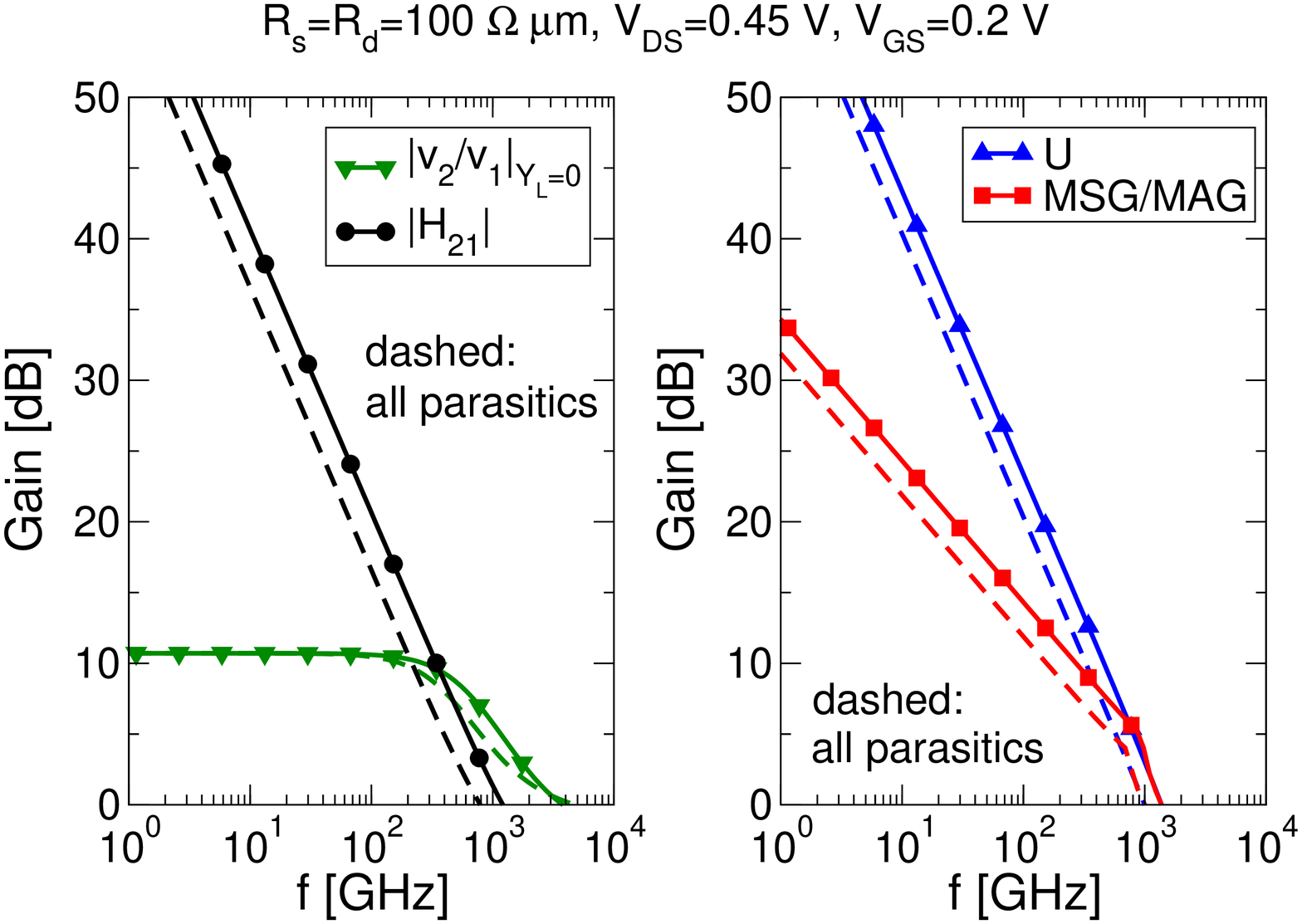}
\caption{Plots of voltage gain $|v_2/v_1|$ with open-circuit load and short-circuit current gain $|H_{21}|$ vs. frequency (left), and of MSG or MAG (depending on frequency) and Mason's unilateral gain $U$ vs. frequency (right) for a $1$-$\mu$m-wide device at $V_{DS}=0.45$~V and $V_{GS}=0.2$~V (quasi-saturation regime). The dashed curves are obtained with $R_g=4$~$\Omega$ and $C_{int}=C_{ext}=0.1$~fF. The other parameters are indicated in the legend.}
\label{fig_gain_vs_f_sat}
\end{figure}
Similar plots of voltage gain, current gain, and power gain, but for the peak-$g_m$ bias point of the quasi-saturation region ($V_{DS}=0.45$~V, $V_{GS}=0.2$~V), are shown in Fig.~\ref{fig_gain_vs_f_sat}. For reference, Mason's unilateral gain $U$ is also included. Here, the voltage gain $|v_2/v_1|$ is computed in the open-circuit-load condition, for which $|A_{v0}|$ takes the maximum value $g_m/g_d\approx 3.4$. It can be noted that the voltage gain bandwidth is significantly wider than in the NDR regime. In the case of $R_g=C_{int}=C_{ext}=0$, values of $\text{GBW}=2.1$~THz, $f_T=1.1$~THz and $f_{max}=2.7$~THz are extracted. The higher $f_T$ in the NDR case can be explained through the beneficial effect of a negative $g_d$ in the denominator of (\ref{eq_ft}), which helps suppress the effect of the source and drain contact resistances. Instead, the lower $f_{max}$ in the NDR regime is caused by the degradation of $g_m$ that was discussed in Section~\ref{sec_dc}. Interestingly, the effect of non-zero $R_g$, $C_{int}$, and $C_{ext}$ is stronger in the quasi-saturation regime than in the NDR regime. This can be explained at least in the case of $C_{ext}=0$ and non-zero $R_g$ and $C_{int}$, for which Eqs.~{\ref{eq_ft}}--{\ref{eq_fmax}} are still valid with $C_{gs}$ and $C_{gd}$ replaced by $C_{gs}+C_{int}$ and $C_{gd}+C_{int}$, respectively: since $g_d$ and $g_m$ have opposite sign, they tend to cancel the contributions of the capacitances. The frequency figures of merit extracted in the different cases are reported in Table~\mbox{\ref{tab_rf}}. All values reported here for $f_{max}$ are more than an order of magnitude higher than the best values measured in fabricated GFETs (40--70~GHz) \cite{Wu12NL,Guo13}, a fact that has to be mainly attributed to the ultra-scaled EOT considered in the simulations (0.5~nm rather than 10--20~nm). It is worth noting that the ideal GFETs considered here compete in terms of $f_{max}$ with III-V HEMTs \cite{Schwierz13}, despite the lack of current saturation.

\begin{table}[!t]
\renewcommand{\arraystretch}{1.3}
\caption{RF metrics at $V_{DS}=0.45$~V and $V_{GS}=0.2$~V (Sat.), and at $V_{DS}=-0.45$~V and $V_{GS}=-1.3$~V (NDR).}
\label{tab_rf}
\centering
\begin{tabular}{c*{4}{c}}
\hline
 & \multicolumn{2}{c}{$R_g=0$ $\Omega$} & \multicolumn{2}{c}{$R_g=4$ $\Omega$} \\
 & \multicolumn{2}{c}{$C_{int}=C_{ext}=0$ fF} & \multicolumn{2}{c}{$C_{int}=C_{ext}=0.1$ fF} \\
\cmidrule(lr){2-5}
& Sat. & NDR & Sat. & NDR \\ 
\hline
GBW [GHz] & 2110 & 195 & 1340 & 172 \\
$f_T$ [GHz] & 1080 & 2280 & 672 & 3020 \\
$f_{max}$ [GHz] $^\textrm{a}$ & 2700 & 891 & 1540 & 935 \\
$f_{max,U}$ [GHz] $^\textrm{b}$ & 1470 & & 1040 & \\
\hline
\vspace{-10pt} \\
\multicolumn{5}{l}{$^\textrm{a}$ Unity frequency of MSG. $^\textrm{b}$ Unity frequency of Mason's unilateral gain.}\\
\end{tabular}
\end{table}

Due to the strong dependence of $|A_{v0}|$ on the ratio $G_L/g_{out}$ (Eq.~\ref{eq_av0}), a strong dependence of $|A_{v0}|$ on the operating point is expected too, leading to undesirable non-linearity.
\begin{figure}[!t]
\centering
\includegraphics[scale=0.32]{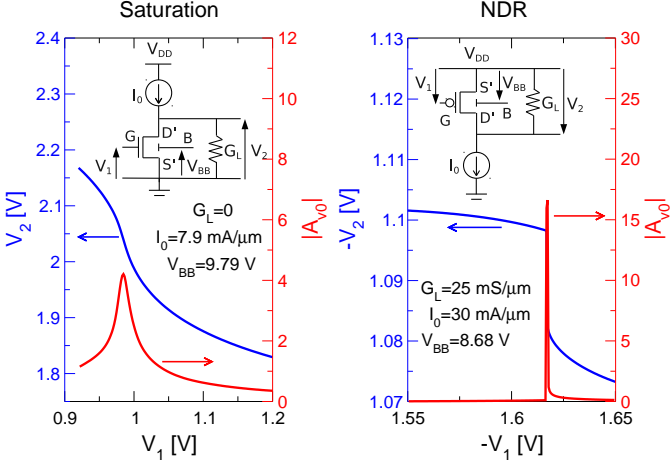}
\caption{Voltage transfer characteristics and corresponding voltage gain characteristics for the device biased in the quasi-saturation (left) and NDR region (right), assuming $R_s=R_d=100$~$\Omega \cdot \mu$m. The biasing circuit is shown in the inset of each figure. The discontinuity of $|A_{v0}|$ in (right) is an artifact of the spline interpolation.}
\label{fig_vin_vout}
\end{figure}
In order to check this, we have considered the biasing circuit shown in the inset of Fig.~\ref{fig_vin_vout}-left and of Fig.~\ref{fig_vin_vout}-right, for the quasi-saturation and the NDR case, respectively. The circuit equations have been solved using a spline interpolation of the $I$--$V$ characteristics with varying $V_{BS}$. The resulting $V_2$--$V_1$ characteristics and the respective voltage gain characteristics are shown in Fig.~\ref{fig_vin_vout}. The higher peak of the voltage gain in the NDR region compared to the quasi-saturation region is obtained at the cost of a much narrower voltage range available to the input signal, which thus limits the use of the device to small-swing signals ($<1$~mV) such as those present at the input of a high-speed pre-amplifier stage. While in the quasi-saturation regime the voltage transfer characteristics are similar to those obtained in graphene complimentary inverter amplifiers \cite{Traversi09,Appenzeller12}, in the NDR case they resemble the typical characteristics of an inverter amplifier with a high-gain region, thus suggesting negative feedback applications. It should be noted, however, that the output voltage swing is quite limited ($<20$~mV).

\section{Conclusions}
\label{sec_conclusions}
In this work, we have investigated the possibility of employing a GFET biased in the region of NDR to achieve higher voltage gains in RF applications. Through a small-signal analysis with parameters extracted from atomistic quantum transport simulations, the stability and RF performance of the transistor in the common-source amplifier configuration have been evaluated. Stability has been found to be a critical issue: compensation of the negative real part of the input and output admittances  is required by means of a careful calibration of the source and load networks. Such compensation can be unfeasible if the series parasitic inductance is too large. Voltage gains exceeding the intrinsic gain in the quasi-saturation regime and larger than 10 can actually be achieved. However, this comes at the expenses of a voltage swing available to the input signal smaller than 1~mV and of a reduced bandwidth. Also, $f_{max}$ is found to be smaller than in the quasi-saturation regime as a result of a four-fold decrease of $g_m$, which is intrinsically related to the device physics responsible for the NDR mechanism.

\appendices


\section*{Acknowledgment}

The authors would like to thank Prof. A.~Santarelli of University of Bologna and Dr. T.~Low of IBM T. J.
Watson Research Center for useful discussions.

\ifCLASSOPTIONcaptionsoff
  \newpage
\fi



\bibliographystyle{IEEEtran}
\bibliography{IEEEabrv,mybibfile}
\end{document}